# Particular and Unique solutions of DGLAP evolution equations in Next-to-Next-to-Leading Order and Structure Functions at low-*x*

Rasna Rajkhowa

Physics Department, T.H.B. College, Jamugurihat, Sonitpur, Assam, India.

E-mail:rasna.rajkhowa@gmail.com

**Abstract**

We present particular and unique solutions of singlet and non-singlet Dokshitzer-Gribov-Lipatov- Altarelli-Parisi (DGLAP) evolution equations in next-to-next-to-leading order (NNLO) at low-x. We obtain *t*-evolutions of deuteron, proton, neutron and difference and ratio of proton and neutron structure functions at low-*x* from DGLAP evolution equations. The results of *t*-evolutions are compared with HERA and NMC lox-*x* and low-$Q^2$ data. We also compare our result of *t*-evolution of proton structure function with a recent global parameterization.

PACS No.: 12.38.Bx, 12.39-x, 13.60Hb.

Keywords: Particular solution, Complete solution, Unique solution, Altarelli-Parisi equation, Structure function, Low-*x* physics.

## 1. Introduction

In our earlier works [1-4], we obtain particular solution of the Dokshitzer-Gribov-Lipatov- Altarelli-Parisi (DGLAP) evolution equations [5-8] for *t* and *x*-evolutions of singlet and non-singlet structure functions in leading order (LO) and next-to-leading order (NLO) and hence *t*-evolution of deuteron, proton, neutron, difference and ratio of proton and neutron and *x*-evolution of deuteron in LO and NLO at low-*x* have been reported. The same technique can be applied to the DGLAP evolution equations in next-to-next-to-leading order (NNLO) for singlet and non-singlet structure functions to obtain *t*-evolutions of deuteron, proton, neutron, difference and ratio of proton and neutron structure functions. These NNLO results are compared with the HERA H1 [9] and NMC [10] low-*x*, low-$Q^2$ data and we also compare our results of t-evolution of proton structure functions with recent global parameterization [11].

## 2. Theory

The DGLAP evolution equations with splitting functions for singlet and non-singlet structure functions in NNLO are in the standard forms [12-14]

$$\frac{\partial F_2^S(x,t)}{\partial t} - \frac{\alpha_S(t)}{2\pi}\left[\frac{2}{3}\{3+4\ln(1-x)\}F_2^S(x,t) + I_1^S(x,t)\right] - \left(\frac{\alpha_S(t)}{2\pi}\right)^2 I_2^S(x,t) - \left(\frac{\alpha_S(t)}{2\pi}\right)^3 I_3^S(x,t) = 0 \qquad (1)$$

and

$$\frac{\partial F_2^{NS}(x,t)}{\partial t} - \frac{\alpha_S(t)}{2\pi}\left[\frac{2}{3}\{3+4\ln(1-x)\}F_2^{NS}(x,t) + I_1^{NS}(x,t)\right] - \left(\frac{\alpha_S(t)}{2\pi}\right)^2 I_2^{NS}(x,t) - \left(\frac{\alpha_S(t)}{2\pi}\right)^3 I_3^{NS}(x,t) = 0 \qquad (2)$$

where, $I_1^S(x,t) = \frac{4}{3}\int_x^1 \frac{d\omega}{1-\omega}\left[(1+\omega^2)F_2^S\left(\frac{x}{\omega},t\right) - 2F_2^S(x,t)\right] + N_f\int_x^1 \{\omega^2+(1-\omega)^2\}G\left(\frac{x}{\omega},t\right)d\omega, \qquad (3)$

$$I_2^S = \left[(x-1)F_2^S(x,t)\int_0^1 f(\omega)d\omega + \int_x^1 f(\omega)F_2^S\left(\frac{x}{\omega},t\right)d\omega + \int_x^1 F_{qq}^S(\omega)F_2^S\left(\frac{x}{\omega},t\right)d\omega + \int_x^1 F_{qg}^S(\omega)G\left(\frac{x}{\omega},t\right)d\omega\right] \qquad (4)$$

$$I_3^S(x,t) = \int_x^1 \frac{d\omega}{\omega}\left[P_{qq}(x)F_2^{NS}\left(\frac{x}{\omega},t\right) + P_{qg}(x)G\left(\frac{x}{\omega},t\right)\right]. \qquad (5)$$

$$I_1^{NS}(x,t) = \frac{4}{3}\int_x^1 \frac{d\omega}{1-\omega}\left[(1+\omega^2)F_2^{NS}\left(\frac{x}{\omega},t\right) - 2F_2^{NS}(x,t)\right], \qquad (6)$$

$$I_2^{NS}(x,t) = (x-1)F_2^{NS}(x,t)\int_0^1 f(\omega)d\omega + \int_x^1 f(\omega)F_2^{NS}\left(\frac{x}{\omega},t\right)d\omega \qquad (7)$$

$$I_3^{NS}(x,t) = \int_x^1 \frac{d\omega}{\omega}\left[P_{NS}^{(2)}(x)F_2^{NS}\left(\frac{x}{\omega},t\right)\right]. \qquad (8)$$

Here $t = \ln\frac{Q^2}{\Lambda^2}$, $\Lambda$ is the QCD cut off parameter. Also

$$f(\omega) = C_F^2[P_F(\omega) - P_A(\omega)] + \frac{1}{2}C_F C_A[P_G(\omega) + P_A(\omega)] + C_F T_R N_f P_{N_f}(\omega),$$

$$F_{qq}^S(\omega) = 2C_F T_T N_f F_{qq}(\omega),\; F_{qg}^S(\omega) = C_F T_T N_f F_{qg}^1(\omega) + C_G T_T N_f F_{qg}^2(\omega)$$

$$P_F(\omega) = -\frac{2(1+\omega^2)}{(1-\omega)}\ln(\omega)\ln(1-\omega) - \left(\frac{3}{1-\omega} + 2\omega\right)\ln\omega - \frac{1}{2}(1+\omega)\ln\omega + \frac{40}{3}(1-\omega),$$

$$P_G(\omega) = \frac{(1+\omega^2)}{(1-\omega)}\left(\ln^2\omega + \frac{11}{3}\ln\omega + \frac{67}{9} - \frac{\pi^2}{3}\right) + 2(1+\omega)\ln\omega + \frac{40}{3}(1-\omega),$$

$$P_{N_f}(\omega) = \frac{2}{3}\left[\frac{1+\omega^2}{1-\omega}\left(-\ln\omega - \frac{5}{3}\right) - 2(1-\omega)\right]$$

$$P_A(\omega) = \frac{21+\omega^2}{(1+\omega)}\int_{\left(\frac{\omega}{1+\omega}\right)}^{\left(\frac{1}{1+\omega}\right)} \frac{dk}{k}\ln\left(\frac{1-k}{k}\right) + 2(1+\omega)\ln\omega + 4(1-\omega)$$

$$F_{qq}(\omega) = \frac{20}{9\omega} - 2 + 6\omega - \frac{56}{9}\omega^2 + \left(1 + 5\omega + \frac{8}{3}\omega^2\right)\ln\omega - (1+\omega)\ln^2\omega$$

$$F_{qg}^{1}(\omega)=4-9\omega-(1-4\omega)\ln\omega-(1-2\omega)\ln^{2}\omega+4\ln(1-\omega)+\left[2\ln^{2}\left(\frac{1-\omega}{\omega}\right)-4\ln\left(\frac{1-\omega}{\omega}\right)-\frac{2}{3}\pi^{2}+10\right]P_{qg}^{1}(\omega)$$

$$F_{qg}^{2}(\omega)=\frac{182}{9}+\frac{14}{9}\omega+\frac{40}{9\omega}+\left(\frac{136}{3}\omega-\frac{38}{3}\right)\ln\omega-4\ln(1-\omega)-(2+8\omega)\ln^{2}\omega$$

$$+\left[-\ln^{2}\omega+\frac{44}{3}\ln\omega-2\ln^{2}(1-\omega)+4\ln(1-\omega)+\frac{\pi^{2}}{3}-\frac{218}{3}\right]P_{qg}^{1}(\omega)+2P_{qg}^{1}(-\omega)\int_{\left(\frac{\omega}{1+\omega}\right)}^{\left(\frac{1}{1+\omega}\right)}\frac{dz}{z}\ln\frac{1-z}{z}$$

$P_{qg}^{1}(\omega)=\omega^{2}+(1-\omega)^{2}$, $N_c = C_A = C_G = 3$, $T_R = ½$, $C_F(\omega)=\dfrac{(N_c^{2}-1)}{2N_c}$, $P_{qq}=P_{NS}^{(2)}+P_{PS}^{(2)}$

$$P_{NS}^{(2)}(x,t)=N_f\left\{\begin{array}{l}\{L_1(-163.9x^{-1}-7.208x)+151.49+44.51x-43.12x^{2}+4.82x^{3}\}(1-x)\\+L_0L_1(-173.1+46.18L_0)+178.04L_0+6.892L_0^{2}+\frac{40}{27}(L_0^{4}-2L_0^{3})\end{array}\right\}$$

$$P_{PS}^{(2)}(x)\cong\left\{\begin{array}{l}N_f\left(\begin{array}{l}-5.926L_1^{3}-9.751L_1^{2}-72.11L_1+177.4+392.9x-101.4x^{2}\\-57.04L_0L_1-661.6L_0+131.4L_0^{2}-\frac{400}{9}L_0^{3}+\frac{160}{27}L_0^{4}\\-506.0x^{-1}-\frac{3584}{27}x^{-1}L_0\end{array}\right)\\+N_f^{2}\left(\begin{array}{l}1.778L_1^{2}+5.944L_1+100.1-125.2x+49.26x^{2}-12.59x^{3}\\-1.889L_0L_1+61.75L_0+17.89L_0^{2}+\frac{32}{27}L_0^{3}+\frac{256}{81}x^{-1}\end{array}\right)\end{array}\right\}(1-x)$$

$$P_{qg}(x)\cong N_f\left(\begin{array}{l}\frac{100}{27}L_1^{4}-\frac{70}{9}L_1^{3}-120.5L_1^{2}+104.42L_1+2522-3316x+2126x^{2}\\+L_0L_1(1823-25.22L_0)-252.5xL_0^{3}+424.9L_0+881.5L_0^{2}\\-\frac{44}{3}L_0^{3}+\frac{536}{27}L_0^{4}-1268.3x^{-1}-\frac{896}{3}x^{-1}L_0\end{array}\right)$$

$$+N_f^{2}\left(\begin{array}{l}\frac{20}{27}L_1^{3}+\frac{200}{27}L_1^{2}-5.496L_1-252.0+158.0x+145.4x^{2}-139.28x^{3}\\-98.07xL_0^{2}+11.70xL_0^{3}-L_0L_1(53.09+80.616L_0)\\-254.0L_0-90.80L_0^{2}-\frac{376}{27}L_0^{3}-\frac{16}{9}L_0^{4}+\frac{1112}{243}x^{-1}\end{array}\right)$$

with $L_0 = \ln(x)$, $L_1 = \ln(1-x)$.

Here results are from direct x-space evolution and $P_{NS}^{(2)}(x)$ is calculated using Fortran package [15]. Except for $x$ values very close to zero of $P_{NS}^{(2)}(x)$, this parameterizations deviate from the exact expressions by less than one part in thousand, which can be consider as sufficiently accurate. For a maximal accuracy for the convolutions with quark densities, slight adjustment should done using low integer moments [16].

The strong coupling constant, $\alpha_S(Q^2)$ is related with the $\beta$-function as [17]

$$\beta(\alpha_S) = \frac{\partial \alpha_S(Q^2)}{\partial \log Q^2} = -\frac{\beta_0}{4\pi}\alpha_s^2 - \frac{\beta_1}{16\pi^2}\alpha_s^3 - \frac{\beta_2}{64\pi^3}\alpha_S^4 + \cdots.$$

where

$$\beta_0 = \frac{11}{3}N_C - \frac{4}{3}T_f = \frac{33 - 2N_f}{3},$$

$$\beta_1 = \frac{34}{3}N_C^2 - \frac{10}{3}N_C N_f - 2C_F N_f = \frac{306 - 38N_f}{3}$$

$$\beta_2 = \frac{2857}{54}N_C^3 + 2C_F^2 T_f - \frac{205}{9}C_F N_C T_f + \frac{44}{9}C_F T_f^2 + \frac{158}{27}N_C T_f^2 = \frac{2857}{2} - \frac{6673}{18}N_f + \frac{325}{54}N_f^2$$

Let us now introduce the variable $u$ = 1-$w$ and note that [18]

$$\frac{x}{w} = \frac{x}{1-u} = x\sum_{k=0}^{\infty} u^k.$$

The above series is convergent for $|u|<1$. Since $x<w<1$, so $0<u<1-x$ and hence the convergence criterion is satisfied. Now, using Taylor expansion method we can rewrite $F_2^S(x/w, t)$ as

$$F_2^S(x/w,t) = F_2^S(x + x\sum_{k=1}^{\infty} u^k, t) = F_2^S(x,t) + x\sum_{k=1}^{\infty} u^k \frac{\partial F_2^S(x,t)}{\partial x} + \frac{1}{2}x^2(\sum_{k=1}^{\infty} u^k)^2 \frac{\partial^2 F_2^S(x,t)}{\partial x^2} + .....$$

which covers the whole range of $u$, $0<u<1-x$. Since $x$ is small in our region of discussion, the terms containing $x^2$ and higher powers of $x$ can be neglected as our first approximation as discussed in our earlier works [1-4], $F_2^S(x/w, t)$ can be approximated for small-$x$ as

$$F_2^S(x/w,t) \cong F_2^S(x,t) + x\sum_{k=1}^{\infty} u^k \frac{\partial F_2^S(x,t)}{\partial x}. \qquad (9)$$

Similarly, $G(x/w, t)$ and $F_2^{NS}(x/w, t)$ can be approximated for small-$x$ as

$$G(x/w,t) \cong G(x,t) + x\sum_{k=1}^{\infty} u^k \frac{\partial G(x,t)}{\partial x} \qquad (10)$$

and

$$F_2^{NS}(x/w,t) \cong F_2^{NS}(x,t) + x\sum_{k=1}^{\infty} u^k \frac{\partial F_2^{NS}(x,t)}{\partial x}. \qquad (11)$$

Using equations (9) and (10) in equations (3), (4) and (5) and performing $u$-integrations we get,

$$\frac{\partial F_2^S(x,t)}{\partial t}-\frac{\alpha_S(t)}{2\pi}\left[A_1(x)F_2^S(x,t)+A_2(x)\frac{\partial F_2^S(x,t)}{\partial x}+A_3(x)G(x,t)+A_4(x)\frac{\partial G(x,t)}{\partial x}\right]$$

$$-\left(\frac{\alpha_S(t)}{2\pi}\right)^2\left[B_1(x)F_2^S(x,t)+B_2(x)\frac{\partial F_2^S(x,t)}{\partial x}+B_3(x)G(x,t)+B_4(x)\frac{\partial G(x,t)}{\partial x}\right]$$

$$-\left(\frac{\alpha_S(t)}{2\pi}\right)^3\left[C_1(x)F_2^S(x,t)+C_2(x)\frac{\partial F_2^S(x,t)}{\partial x}+C_3(x)G(x,t)+C_4(x)\frac{\partial G(x,t)}{\partial x}\right]=0 \qquad (12)$$

where,

$$A_1(x)=2x+x^2+4\ln(1-x), \qquad A_2(x)=x-x^3-2x\ln x,$$

$$A_3(x)=2N_f\left(\frac{2}{3}-x+x^2-\frac{2}{3}x^3\right), \qquad A_4(x)=2N_f\left(-\frac{5}{3}x+3x^2-2x^3+\frac{2}{3}x^4-x\ln x\right),$$

$$B_1(x)=x\int_0^1 f(\omega)d\omega-\int_0^x f(\omega)d\omega+\frac{4}{3}N_f\int_x^1 F_{qq}(\omega)d\omega,$$

$$B_2(x)=x\int_x^1\left[f(\omega)+\frac{4}{3}N_f F_{qg}^S(\omega)\right]\frac{1-\omega}{\omega}d\omega,$$

$$B_3(x)=\int_x^1 F_{qg}^S(\omega)d\omega, \qquad B_4(x)=x\int_x^1\frac{1-\omega}{\omega}F_{qg}^S(\omega)d\omega$$

$$C_1(x)=N_f\int_0^{1-x}\frac{d\omega}{1-\omega}C(\omega), \qquad C(x)=N_f\int_0^{1-x}\frac{\omega x dx}{(1-\omega)^2}C(\omega),$$

$$C_3(x)=N_f\int_0^{1-x}\frac{d\omega}{1-\omega}C^{/}(\omega), \qquad C_2(x)=N_f\int_0^{1-x}\frac{\omega x dx}{(1-\omega)^2}C^{/}(\omega),$$

here,

$$C(\omega)=\left[\begin{array}{l}\ln\omega\ln(1-\omega)[-173.1+46.18\ln(1-\omega)]+178.04\ln(1-\omega)+6.892\ln^2(1-\omega)\\+\frac{40}{27}(\ln^4(1-\omega)-2\ln^3(1-\omega))\end{array}\right]$$

$$+\omega\left\{\begin{array}{l}\ln\omega[-163.9(1-\omega)^{-1}-7.208(1-\omega)]+151.49+44.51(1-\omega)-43.12(1-\omega)^2+4.82(1-\omega)^3-5.926\ln^3\omega\\-9.751\ln^2\omega-72.11\ln\omega+177.4+392.9(1-\omega)-101.4(1-\omega)^2-57.04\ln(1-\omega)\ln\omega-661.6\ln(1-\omega)\\+131.4\ln^2(1-\omega)-\frac{400}{9}\ln^3(1-\omega)+\frac{160}{27}\ln^4(1-\omega)-506(1-\omega)^{-1}-\frac{3584}{27}(1-\omega)^{-1}\ln(1-\omega)\end{array}\right\}$$

$$+N_f\omega\begin{bmatrix}1.778\ln^2\omega+5.944\ln\omega+100.1-125.2(1-\omega)+49.26(1-\omega)^2-12.59(1-\omega)^3-1.889\ln(1-\omega)\ln\omega\\+61.75\ln(1-\omega)+17.89\ln^2(1-\omega)+\frac{32}{27}\ln^3(1-\omega)+\frac{256}{81}(1-\omega)^{-1}\end{bmatrix}$$

$$C'(\omega)=\begin{Bmatrix}\frac{100}{27}\ln^4\ln\omega-\frac{70}{9}\ln^3\omega-120.5\ln^2\omega+104.42\ln\omega+2522\text{-}3316(1\text{-}\omega)+2126(1\text{-}\omega)^2\\-252.5(1-\omega)\ln^3(1-\omega)+\ln\omega\ln(1-\omega)(1823-25.22\ln(1-\omega))+424.9\ln(1-\omega)\\+881.5\ln^2(1-\omega)-\frac{44}{3}\ln^3\omega+\frac{536}{27}\ln^4(1-\omega)-1268.3(1-\omega)^{-1}-\frac{896}{3}(1-\omega)^{-1}\ln(1-\omega)\end{Bmatrix}$$

$$+N_f\begin{Bmatrix}\frac{20}{27}\ln^3\omega+\frac{200}{27}\ln^2\omega-5.496\ln\omega-252+158(1-\omega)+145.4(1-\omega)^2-139.28(1-\omega)^3\\-98.07(1-\omega)\ln^2(1-\omega)+11.70(1-\omega)\ln^3(1-\omega)-\ln\omega\ln(1-\omega)(53.09+80.616\ln(1-\omega))\\-254\ln(1-\omega)-90.80\ln^2(1-\omega)-\frac{376}{27}\ln^3(1-\omega)-\frac{16}{9}\ln^4(1-\omega)+\frac{1112}{243}(1-\omega)^{-1}\end{Bmatrix}$$

Let us assume for simplicity [1-4]

$$G(x,t)=K(x)\,F_2^S(x,t) \tag{13}$$

where $K(x)$ is a function of $x$. In this connection, earlier we considered [1-3] $K(x) = k, ax^b, ce^{-dx}$, where $k, a, b, c, d$ are constants. Agreement of the results with experimental data is found to be excellent for $k = 4.5$, $a = 4.5$, $b = 0.01$, $c = 5$, $d = 1$ for low-$x$ in leading order and $a = 10$, $b = 0.016$, $c = 0.5$, $d = -3.8$ in next-to-leading order. Also we can consider two numerical parameters $T_0$ and $T_1$ such that

$$T^2(t) = T_0 T(t) \text{ and } T^3(t) = T_0 T(t)\, T(t) = T_1 T(t), \tag{14}$$

where, $T(t)=\dfrac{\alpha_S(t)}{2\pi}$

By suitable choice of $T_0$ and $T_1$, we can reduce the error to a minimum. Hence equation (12) becomes

$$\frac{\partial F_2^S(x,t)}{\partial t}-P_S(x)\frac{\partial F_2^S(x,t)}{\partial x}-Q_S(x)F_2^S(x,t)=0, \tag{15}$$

where,

$$P_S(x)=T(t)[(A_2+KA_4)+T_0(B_2+KB_4)+T_1(C_2+KC_4)],$$

$$Q_S(x)=T(t)\left[(A_1+KA_3+A_4\frac{\partial K}{\partial x})+T_0(B_1+KB_3+B_4\frac{\partial K}{\partial x})+T_1(C_1+KC_3+C_4\frac{\partial K}{\partial x})\right]$$

Secondly, using equations (11) and (14) in equations (6), (7) and (8) and performing $u$-integration equation (2) becomes

$$\frac{\partial F_2^{NS}(x,t)}{\partial t} - P_{NS}(x)\frac{\partial F_2^{NS}(x,t)}{\partial x} - Q_{NS}(x)F_2^{NS}(x,t) = 0 \qquad (16)$$

where,

$P_{NS}(x) = T(t)[A_2 + T_0 B_2 + T_1 C_2]$,

$Q_S(x) = T(t)[A_1 + T_0 B_1 + T_1 C_1]$

The general solutions [19, 20] of equations (15) is $F(U, V) = 0$, where $F$ is an arbitrary function and $U(x, t, F_2^S) = C_1$ and $V(x, t, F_2^S) = C_2$ where, $C_1$ and $C_2$ are constant and they form a solution of equations

$$\frac{dx}{P_S(x,t)} = \frac{dt}{-1} = \frac{dF_2^S(x,t)}{-Q_S(x,t)}. \qquad (17)$$

We observed that the Lagrange's auxiliary system of ordinary differential equations [19, 20] occurred in the formalism can not be solved without the additional assumption of linearization (equation (14)) and introduction two numerical parameters $T_0$ and $T_1$. These parameters does not effect in the results of t- evolution of structure functions. Solving equation (17) we obtain,

$$U(x,t,F_2^S) = t^{\left(1+\frac{b}{t}-\frac{c}{2t^2}\ln t\right)} \exp\left[\frac{N_S(x)}{a} + \frac{b}{t} + \frac{1}{t^2}\left(\frac{c}{2} - \frac{d}{2}\right)\right] \text{ and}$$

$$V(x,t,F_2^S) = F_2^S(x,t)\exp[M_S(x)]$$

where,

$$a = \frac{2}{\beta_0}, \qquad b = \frac{\beta_1}{\beta_0^2}, \qquad c = \frac{\beta_1}{\beta_0^4}, \qquad d = \frac{\beta_2}{\beta_0^3},$$

$$N_S(x) = \int \frac{dx}{[(A_2 + KA_4) + T_0(B_2 + KB_4) + T_1(C_2 + KC_4)]}$$

and

$$M_S(x) = \int \frac{(A_1 + KA_3 + A_4\frac{\partial K}{\partial x}) + T_0(B_1 + KB_3 + B_4\frac{\partial K}{\partial x}) + T_1(C_1 + KC_3 + C_4\frac{\partial K}{\partial x})}{(A_2 + KA_4) + T_0(B_2 + KB_4) + T_1(C_2 + KC_4)} dx.$$

**2. (a) Particular Solutions**

If $U$ and $V$ are two independent solutions of equation (17) and if $\alpha$ and $\beta$ are arbitrary constants, then $V = \alpha U + \beta$ may be taken as a complete solution of equation (15). Then the complete solution [19, 20]

$$F_2^S(x,t)\exp[M_S(x)] = \alpha t^{\left(1+\frac{b}{t}-\frac{c}{2t^2}\ln t\right)} \exp\left[\frac{N_S(x)}{a}+\frac{b}{t}+\frac{1}{t^2}\left(\frac{c}{2}-\frac{d}{2}\right)\right]+\beta \tag{18}$$

is a two-parameter family of planes. The one parameter family determined by taking $\beta = \alpha^2$ has equation

$$F_2^S(x,t)\exp[M_S(x)] = \alpha t^{\left(1+\frac{b}{t}-\frac{c}{2t^2}\ln t\right)} \exp\left[\frac{N_S(x)}{a}+\frac{b}{t}+\frac{1}{t^2}\left(\frac{c}{2}-\frac{d}{2}\right)\right]+\alpha^2. \tag{19}$$

Differentiating equation (19) with respect to $\alpha$, we obtain

$$\alpha = -\frac{1}{2} t^{\left(1+\frac{b}{t}-\frac{c}{2t^2}\ln t\right)} \exp\left[\frac{N_S(x)}{a}+\frac{b}{t}+\frac{1}{t^2}\left(\frac{c}{2}-\frac{d}{2}\right)\right]$$

Putting the value of $\alpha$ again in equation (19), we obtain envelope

$$F_2^S(x,t)\exp[M_S(x)] = -\frac{1}{4}\left[t^{\left(1+\frac{b}{t}-\frac{c}{2t^2}\ln t\right)} \exp\left(\frac{N_S(x)}{a}+\frac{b}{t}+\frac{1}{t^2}\left(\frac{c}{2}-\frac{d}{2}\right)\right)\right]^2.$$

Therefore,

$$F_2^S(x,t) = -\frac{1}{4}\left[t^{2\left(1+\frac{b}{t}-\frac{c}{2t^2}\ln t\right)} \exp\left(\frac{2b}{t}+\frac{2}{t^2}\left(\frac{c}{2}-\frac{d}{2}\right)+\frac{2N_S(x)}{a}-M_S(x)\right)\right] \tag{20}$$

which is merely a particular solution of the general solution. Now, defining

$$F_2^S(x,t_0) = -\frac{1}{4}\left[t_0^{2\left(1+\frac{b}{t_0}-\frac{c}{2t_0^2}\ln t_0\right)} \exp\left(\frac{2b}{t_0}+\frac{2}{t_0^2}\left(\frac{c}{2}-\frac{d}{2}\right)+\frac{2N_S(x)}{a}-M_S(x)\right)\right].$$

at $t = t_0$, where, $t_0 = \ln(Q_0^2/\Lambda^2)$ at any lower value $Q = Q_0$, we get from equation (20)

$$F_2^S(x,t) = F_2^S(x,t_0)\left(\frac{t^{(1+b/t-\frac{c}{2t^2}\ln t)}}{t_0^{(1+b/t_0-\frac{c}{2t_0^2}\ln t_0)}}\right)^2 \exp\left[2b\left(\frac{1}{t}-\frac{1}{t_0}\right)+2\left(\frac{1}{t^2}-\frac{1}{t_0^2}\right)\left(\frac{c}{2}-\frac{d}{2}\right)\right], \tag{21}$$

which gives the $t$-evolution of singlet structure function $F_2^S(x,t)$ in NNLO for $\beta = \alpha^2$.

Proceeding exactly in the same way, and defining

$$F_2^{NS}(x,t_0) = -\frac{1}{4}\left[t_0^{\,2\left(1+\frac{b}{t_0}-\frac{c}{2t_0^2}\ln t\right)}\exp\left(\frac{2b}{t_0}+\frac{2}{t_0^2}\left(\frac{c}{2}-\frac{d}{2}\right)+\frac{2N_{NS}(x)}{a}-M_{NS}(x)\right)\right]$$

where, $N_{NS}(x) = \int \frac{dx}{A_2 + T_0 B_2 + T_1 C_2}$ and $M_{NS}(x) = \int \frac{A_1 + T_0 B_1 + T_1 C_1}{A_2 + T_0 B_2 + T_1 C_2} dx,$

we get,

$$F_2^{NS}(x,t) = F_2^{NS}(x,t_0)\left(\frac{t^{(1+b/t-\frac{c}{2t^2}\ln t)}}{t_0^{(1+b/t_0-\frac{c}{2t_0^2}\ln t_0)}}\right)^2 \exp\left[2b\left(\frac{1}{t}-\frac{1}{t_0}\right)+2\left(\frac{1}{t^2}-\frac{1}{t_0^2}\right)\left(\frac{c}{2}-\frac{d}{2}\right)\right], \tag{22}$$

which gives the $t$-evolution of non-singlet structure function $F_2^{NS}(x, t)$ in NNLO for $\beta = \alpha^2$.

We observe that if $b$, $c$ and $d$ tends to zero, then equations (21) and (22) tends to LO equation [1] and if $c$ and $d$ tends to zero then these equations tends to NLO equation [2-3]. Physically $b$, $c$ and $d$ tends to zero means number of flavours is high.

Again defining,

$$F_2^{S}(x_0,t) = -\frac{1}{4}\left[t^{\,2\left(1+\frac{b}{t}-\frac{c}{2t^2}\ln t\right)}\exp\left(\frac{2b}{t}+\frac{2}{t^2}\left(\frac{c}{2}-\frac{d}{2}\right)+\frac{2N_S(x)}{a}-M_S(x)\right)\right]_{x=x_0}.$$

we obtain from equation (20)

$$F_2^{S}(x,t) = F_2^{S}(x_0,t)\exp\int_{x_0}^{x}\left[\frac{2}{a}.N_S(x)-M_S(x)\right]dx, \tag{23}$$

which gives the $x$-evolution of singlet structure function $F_2^{S}(x, t)$ in NNLO for $\beta = \alpha^2$.

Similarly defining,

$$F_2^{NS}(x_0,t) = -\frac{1}{4}\left[t^{\,2\left(1+\frac{b}{t}-\frac{c}{2t^2}\ln t\right)}\exp\left(\frac{2b}{t}+\frac{2}{t^2}\left(\frac{c}{2}-\frac{d}{2}\right)+\frac{2N_{NS}(x)}{a}-M_{NS}(x)\right)\right]_{x=x_0}.$$

We get

$$F_2^{NS}(x,t) = F_2^{NS}(x_0,t)\exp\int_{x_0}^{x}\left[\frac{2}{a}.N_{NS}(x) - M_{NS}(x)\right]dx, \tag{24}$$

which gives the $x$-evolution of non-singlet structure function $F_2^{NS}(x, t)$ in NNLO for $\beta = \alpha^2$.

Deuteron, proton and neutron structure functions measured in deep inelastic electro-production can be written in terms of singlet and non-singlet quark distribution functions [21] as

$$F_2^d(x, t) = 5/9\, F_2^S(x, t), \tag{25}$$

$$F_2^p(x, t) = 5/18\, F_2^S(x, t) + 3/18\, F_2^{NS}(x, t), \tag{26}$$

$$F_2^n(x, t) = 5/18\, F_2^S(x, t) - 3/18\, F_2^{NS}(x, t) \tag{27}$$

and

$$F_2^p(x, t) - F_2^n(x, t) = 1/3\, F_2^{NS}(x, t). \tag{28}$$

Now using equations (21) and (23) in equation (25) we will get $t$ and $x$-evolution of deuteron structure function $F_2^d(x, t)$ at low-$x$ in NNLO as

$$F_2^d(x,t) = F_2^d(x,t_0)\left(\frac{t^{(1+b/t-\frac{c}{2t^2}\ln t)}}{t_0^{(1+b/t_0-\frac{c}{2t_0^2}\ln t_0)}}\right)^2 \exp\left[2b\left(\frac{1}{t}-\frac{1}{t_0}\right)+2\left(\frac{1}{t^2}-\frac{1}{t_0^2}\right)\left(\frac{c}{2}-\frac{d}{2}\right)\right], \tag{29}$$

and

$$F_2^d(x,t) = F_2^d(x_0,t)\exp\int_{x_0}^{x}\left[\frac{2}{a}.N_S(x) - M_S(x)\right]dx, \tag{30}$$

where, the input functions are

$$F_2^d(x,t_0) = \frac{5}{9}F_2^S(x,t_0) \quad \text{and} \quad F_2^d(x_0,t) = \frac{5}{9}F_2^S(x_0,t).$$

Similarly using equations (21) and (22) in equations (26), (27) and (28) we get the $t$ – evolutions of proton, neutron, and difference and ratio of proton and neutron structure functions at low-x in NNLO as

$$F_2^p(x,t) = F_2^p(x,t_0)\left(\frac{t^{(1+b/t-\frac{c}{2t^2}\ln t)}}{t_0^{(1+b/t_0-\frac{c}{2t_0^2}\ln t_0)}}\right)^2 \exp\left[2b\left(\frac{1}{t}-\frac{1}{t_0}\right)+2\left(\frac{1}{t^2}-\frac{1}{t_0^2}\right)\left(\frac{c}{2}-\frac{d}{2}\right)\right], \tag{31}$$

$$F_2^n(x,t) = F_2^n(x,t_0)\left(\frac{t^{(1+b/t-\frac{c}{2t^2}\ln t)}}{t_0^{(1+b/t_0-\frac{c}{2t_0^2}\ln t_0)}}\right)^2 \exp\left[2b\left(\frac{1}{t}-\frac{1}{t_0}\right)+2\left(\frac{1}{t^2}-\frac{1}{t_0^2}\right)\left(\frac{c}{2}-\frac{d}{2}\right)\right], \quad (32)$$

$$F_2^p(x,t) - F_2^n(x,t) = [F_2^p(x,t_0) - F_2^p(x,t_0)]\left(\frac{t^{(1+b/t-\frac{c}{2t^2}\ln t)}}{t_0^{(1+b/t_0-\frac{c}{2t_0^2}\ln t_0)}}\right)^2 \exp\left[2\left\{b\left(\frac{1}{t}-\frac{1}{t_0}\right)+\left(\frac{1}{t^2}-\frac{1}{t_0^2}\right)\left(\frac{c}{2}-\frac{d}{2}\right)\right\}\right],$$

(33)

and

$$\frac{F_2^p(x,t)}{F_2^n(x,t)} = \frac{F_2^p(x,t_0)}{F_2^n(x,t_0)} = R(x), \quad (34)$$

where R(x) is a constant for fixed-x. And the input functions are

$$F_2^p(x,t_0) = \frac{5}{18}F_2^S(x,t_0) + \frac{3}{18}F_2^{NS}(x,t_0),$$

$$F_2^n(x,t_0) = \frac{5}{18}F_2^S(x,t_0) - \frac{3}{18}F_2^{NS}(x,t_0),$$

and

$$F_2^p(x,t_0) - F_2^n(x,t_0) = \frac{1}{3}F_2^{NS}(x,t_0).$$

For the complete solution of equation (15), we take $\beta = \alpha^2$ in equation (18). We observed that if we take $\beta = \alpha$ in equation (18) and differentiate with respect to $\alpha$ as before, we can not determine the value of $\alpha$. In general, if we take $\beta = \alpha^y$, we get in the solutions, the powers of $\left(\frac{t^{(1+b/t-\frac{c}{2t^2}\ln t)}}{t_0^{(1+b/t_0-\frac{c}{2t_0^2}\ln t_0)}}\right)$ and the co-efficient of $\left[b\left(\frac{1}{t}-\frac{1}{t_0}\right)+\left(\frac{c}{2}-\frac{d}{2}\right)\left(\frac{1}{t^2}-\frac{1}{t_0^2}\right)\right]$ in exponential part in $t$-evolutions of deuteron, proton, neutron, and difference of proton and neutron structure functions be $y/(y-1)$ and the numerators of the first term inside the integral sign be $y/(y-1)$ for $x$- evolutions in NNLO. Hence if $y$ varies from minimum (=2) to maximum (= ∞) then $y/(y-1)$ varies from 2 to 1.

Thus by this methodology, instead of having a single solution we arrive a band of solutions, of course the range for these solutions is reasonably narrow.

## 2. (b) Unique Solutions

Due to conservation of the electromagnetic current, $F_2$ must vanish as $Q^2$ goes to zero [21, 22]. Also $R\rightarrow 0$ in this limit. Here $R$ indicates ratio of longitudinal and transverse cross-sections of virtual photon in DIS process. This implies that scaling should not be a valid concept in the region of very low-$Q^2$. The exchanged photon is then almost real and the close similarity of real photonic and hadronic interactions justifies the use of the Vector Meson Dominance (VMD) concept [23-24] for the description of $F_2$. In the language of perturbation theory, this concept is equivalent to a statement that a physical photon spends part of its time as a 'bare', point-like photon and part as a virtual hadron [22]. The power and beauty of explaining scaling violations with field theoretic methods (i.e., radiative corrections in QCD) remains, however, unchallenged in as much as they provide us with a framework for the whole $x$-region with essentially only one free parameter $\Lambda$ [25]. For $Q^2$ values much larger than $\Lambda^2$, the effective coupling is small and a perturbative description in terms of quarks and gluons interacting weakly makes sense. For $Q^2$ of order $\Lambda^2$, the effective coupling is infinite and we cannot make such a picture, since quarks and gluons will arrange themselves into strongly bound clusters, namely, hadrons [21] and so the perturbation series breaks down at small-$Q^2$ [21]. Thus, it can be thought of $\Lambda$ as marking the boundary between a world of quasi-free quarks and gluons, and the world of pions, protons, and so on. The value of $\Lambda$ is not predicted by the theory; it is a free parameter to be determined from experiment. It should expect that it is of the order of a typical hadronic mass [21]. Since the value of $\Lambda$ is so small we can take at $Q = \Lambda$, $F_2^S(x, t) = 0$ due to conservation of the electromagnetic current [22]. This dynamical prediction agrees with most ad hoc parameterizations and with the data [25]. Using this boundary condition in equation (18) we get $\beta = 0$ and

$$F_2^S(x,t) = \alpha t^{\left(1+\frac{b}{t}-\frac{c}{2t^2}\ln t\right)} \exp\left[\frac{N_S(x)}{a}+\frac{b}{t}+\frac{1}{t^2}\left(\frac{c}{2}-\frac{d}{2}\right)-M_S(x)\right]. \qquad (35)$$

Now, defining

$$F_2^S(x,t_0) = \alpha\left[t_0^{\left(1+\frac{b}{t_0}-\frac{c}{2t_0^2}\ln t_0\right)} \exp\left(\frac{b}{t_0}+\frac{1}{t_0^2}\left(\frac{c}{2}-\frac{d}{2}\right)+\frac{N_S(x)}{a}-M_S(x)\right)\right].$$

at $t = t_0$, we get from equation (35)

$$F_2^S(x,t) = F_2^S(x,t_0)\left(\frac{t^{(1+b/t-\frac{c}{2t^2}\ln t)}}{t_0^{(1+b/t_0-\frac{c}{2t_0^2}\ln t_0)}}\right)\exp\left[b\left(\frac{1}{t}-\frac{1}{t_0}\right)+\left(\frac{1}{t^2}-\frac{1}{t_0^2}\right)\left(\frac{c}{2}-\frac{d}{2}\right)\right], \tag{36}$$

which gives the $t$-evolution of singlet structure function $F_2^S(x, t)$ in NNLO.

Proceeding exactly in the same way, we get

$$F_2^{NS}(x,t) = F_2^{NS}(x,t_0)\left(\frac{t^{(1+b/t-\frac{c}{2t^2}\ln t)}}{t_0^{(1+b/t_0-\frac{c}{2t_0^2}\ln t_0)}}\right)\exp\left[b\left(\frac{1}{t}-\frac{1}{t_0}\right)+\left(\frac{1}{t^2}-\frac{1}{t_0^2}\right)\left(\frac{c}{2}-\frac{d}{2}\right)\right], \tag{37}$$

which gives the $t$-evolution of non-singlet structure function $F_2^{NS}(x, t)$ in NNLO.

Again defining,

$$F_2^S(x_0,t) = \alpha\left[t^{\left(1+\frac{b}{t}-\frac{c}{2t^2}\ln t\right)}\exp\left(\frac{b}{t}+\frac{1}{t^2}\left(\frac{c}{2}-\frac{d}{2}\right)+\frac{N_S(x)}{a}-M_S(x)\right)\right]_{x=x_0}.$$

we obtain from equation (35)

$$F_2^S(x,t) = F_2^S(x_0,t)\exp\int_{x_0}^{x}\left[\frac{1}{a}.N_S(x)-M_S(x)\right]dx, \tag{38}$$

which gives the $x$-evolution of singlet structure function $F_2^S(x, t)$ in NNLO.

Similarly,

$$F_2^{NS}(x_0,t) = \alpha\left[t^{\left(1+\frac{b}{t}-\frac{c}{2t^2}\ln t\right)}\exp\left(\frac{b}{t}+\frac{1}{t^2}\left(\frac{c}{2}-\frac{d}{2}\right)+\frac{N_{NS}(x)}{a}-M_{NS}(x)\right)\right]_{x=x_0}.$$

we get

$$F_2^{NS}(x,t) = F_2^{NS}(x_0,t)\exp\int_{x_0}^{x}\left[\frac{1}{a}.N_{NS}(x)-M_{NS}(x)\right]dx, \tag{39}$$

which gives the $x$-evolution of non-singlet structure function $F_2^S(x, t)$ in NNLO.

Therefore corresponding results for $t$-evolution of deuteron, proton, neutron, difference and ratio of proton and neutron structure functions are

$$F_2^d(x,t)=F_2^d(x,t_0)\left(\frac{t^{\left(1+b/t-\frac{c}{2t^2}\ln t\right)}}{t_0^{\left(1+b/t_0-\frac{c}{2t_0^2}\ln t_0\right)}}\right)\exp\left[b\left(\frac{1}{t}-\frac{1}{t_0}\right)+\left(\frac{1}{t^2}-\frac{1}{t_0^2}\right)\left(\frac{c}{2}-\frac{d}{2}\right)\right], \tag{40}$$

$$F_2^p(x,t)=F_2^p(x,t_0)\left(\frac{t^{\left(1+b/t-\frac{c}{2t^2}\ln t\right)}}{t_0^{\left(1+b/t_0-\frac{c}{2t_0^2}\ln t_0\right)}}\right)\exp\left[b\left(\frac{1}{t}-\frac{1}{t_0}\right)+\left(\frac{1}{t^2}-\frac{1}{t_0^2}\right)\left(\frac{c}{2}-\frac{d}{2}\right)\right], \tag{41}$$

$$F_2^n(x,t)=F_2^n(x,t_0)\left(\frac{t^{\left(1+b/t-\frac{c}{2t^2}\ln t\right)}}{t_0^{\left(1+b/t_0-\frac{c}{2t_0^2}\ln t_0\right)}}\right)\exp\left[b\left(\frac{1}{t}-\frac{1}{t_0}\right)+\left(\frac{1}{t^2}-\frac{1}{t_0^2}\right)\left(\frac{c}{2}-\frac{d}{2}\right)\right], \tag{42}$$

$$F_2^p(x,t)-F_2^n(x,t)=[F_2^p(x,t_0)-F_2^p(x,t_0)]\left(\frac{t^{\left(1+b/t-\frac{c}{2t^2}\ln t\right)}}{t_0^{\left(1+b/t_0-\frac{c}{2t_0^2}\ln t_0\right)}}\right)\exp\left[b\left(\frac{1}{t}-\frac{1}{t_0}\right)+\left(\frac{1}{t^2}-\frac{1}{t_0^2}\right)\left(\frac{c}{2}-\frac{d}{2}\right)\right], \tag{43}$$

and

$$\frac{F_2^p(x,t)}{F_2^n(x,t)}=\frac{F_2^p(x,t_0)}{F_2^n(x,t_0)}=R(x), \tag{44}$$

Again $x$-evolution of deuteron structure function in NNLO is

$$F_2^d(x,t)=F_2^d(x_0,t)\exp\int_{x_0}^{x}\left[\frac{1}{a}.N_S(x)-M_S(x)\right]dx, \tag{45}$$

Already we have mentioned [1-4] that the determination of $x$-evolutions of proton and neutron structure functions like that of deuteron structure function is not suitable by this methodology. It is to be noted that unique solutions of evolution equations of different structure functions are same with particular solutions for $y$ maximum ($y = \infty$) in $\beta = \alpha^y$ relation.

### 3. Results and Discussion

In the present paper, we compare our results of t-evolution of deuteron, proton, neutron and difference and ratio of proton and neutron structure functions with the HERA [9] and NMC [10] low-$x$ and low-$Q^2$ data. In case of HERA data [9] proton and neutron structure functions are measured in the range $2 \leq Q^2 \leq 50$ GeV$^2$. Moreover here $P_T \leq 200$ MeV, where $P_T$ is the transverse momentum of the final state baryon. In case of NMC data proton and deuteron structure functions are measured in the range $0.75 \leq Q^2 \leq 27$GeV$^2$.We consider number of flavours $N_f = 4$. We also compare our results of t-evolution of proton structure functions with recent global parameterization [11]. This parameterization includes data from H1-96 \ 99, ZEUS-96/97(X0.98), NMC, E665 data.

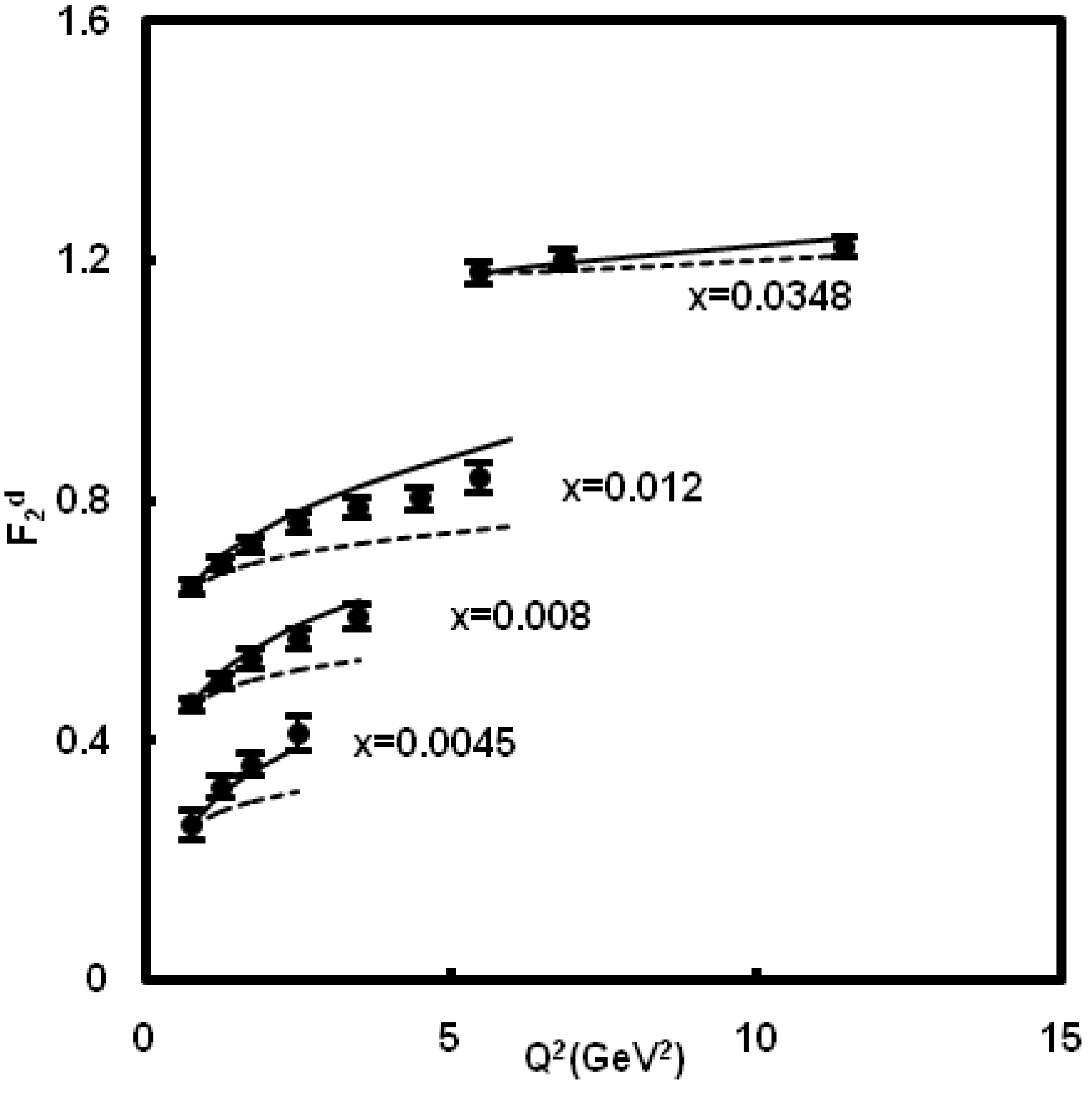

**Fig-1**

In Fig.1, we present our results of t-evolutions of deuteron structure functions for the representative values of $x$ given in the figures for $y = 2$ (solid lines) and $y$ maximum (dashed lines) in $\beta = \alpha^y$ relation. The dashed line also represents results of unique solution. Data points at lowest-$Q^2$ values in the figures are taken as input to test the evolution equation. Agreement with the data [10] is found to be good. NNLO results for $y$ = 2 are of better agreement with experimental data in general.

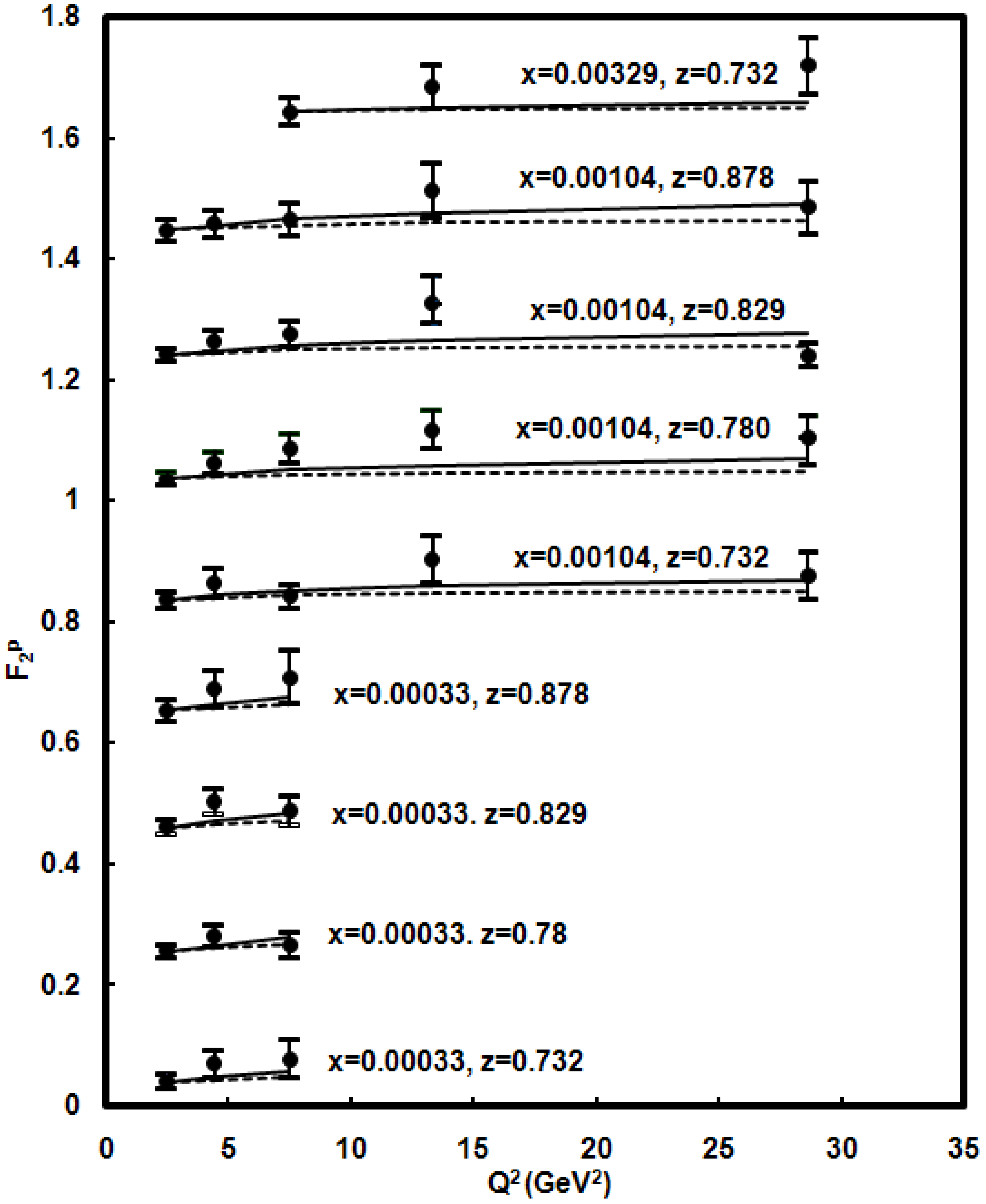

**Fig-2**

In Fig.2, we present our results of t-evolutions of proton structure functions for the representative values of $x$ given in the figures for $y$ = 2 (solid lines) and $y$ maximum (dashed lines) in $\beta = \alpha^y$ relation. The dashed line also represents results of unique solution. Data points at lowest-$Q^2$ values in the figures are taken as input to test the evolution equation. Agreement with the data [9] is found to be good. NNLO results for $y$ = 2 are of better agreement with experimental data in general.

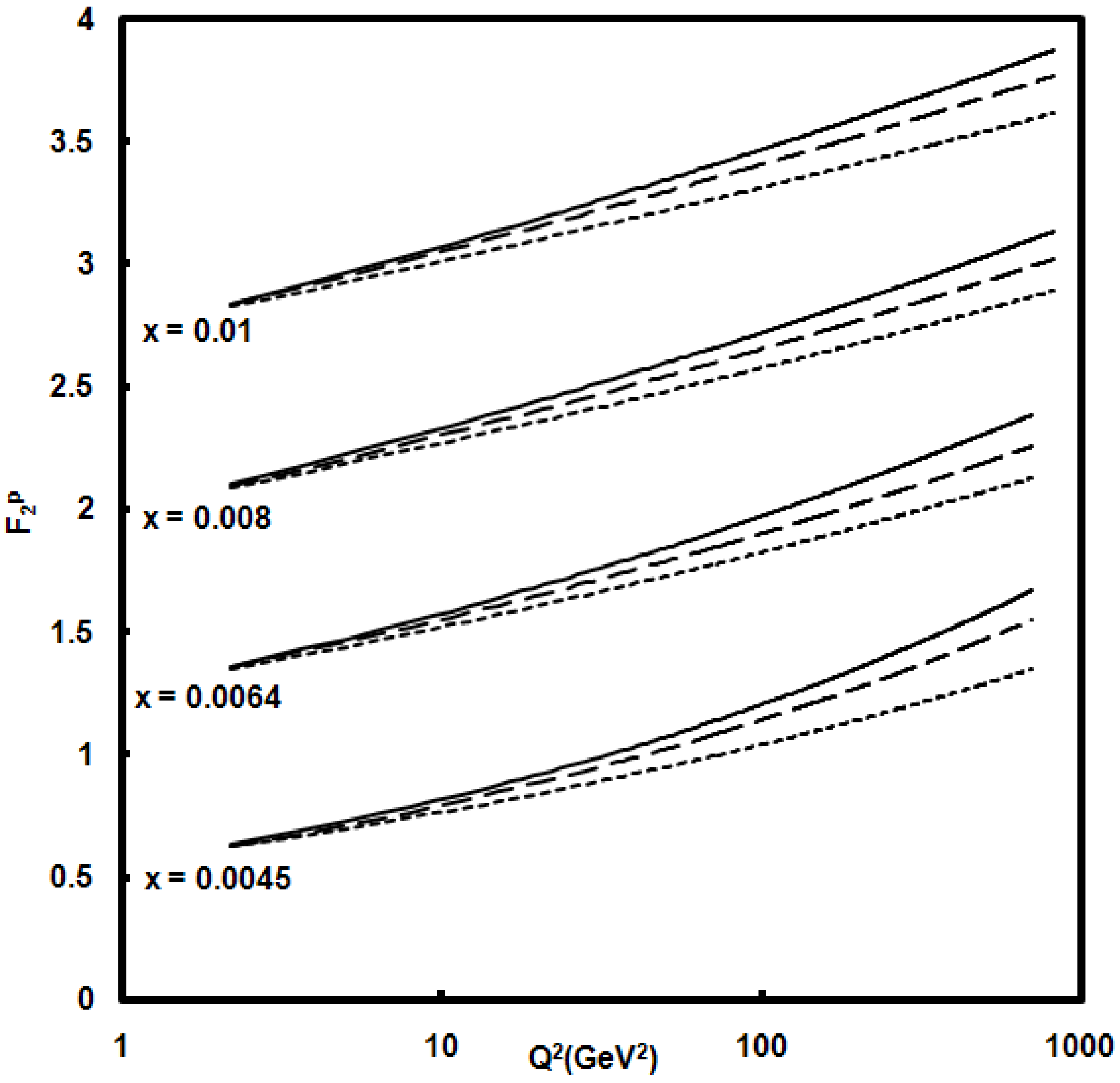

Fig-3

In fig.3 we compare our results of $t$-evolutions of proton structure functions $F_2^p$ with recent global parameterization [11] (long dashed lines) for the representative values of $x$ given in the figures for $y$ = 2 (solid lines) and $y$ maximum (dashed lines) in $\beta = \alpha^y$ relation. The dashed line also represents results of unique solution. Data points at lowest-$Q^2$ values in the figures are taken as input to test the evolution equation. Agreement with the results is found to be good.

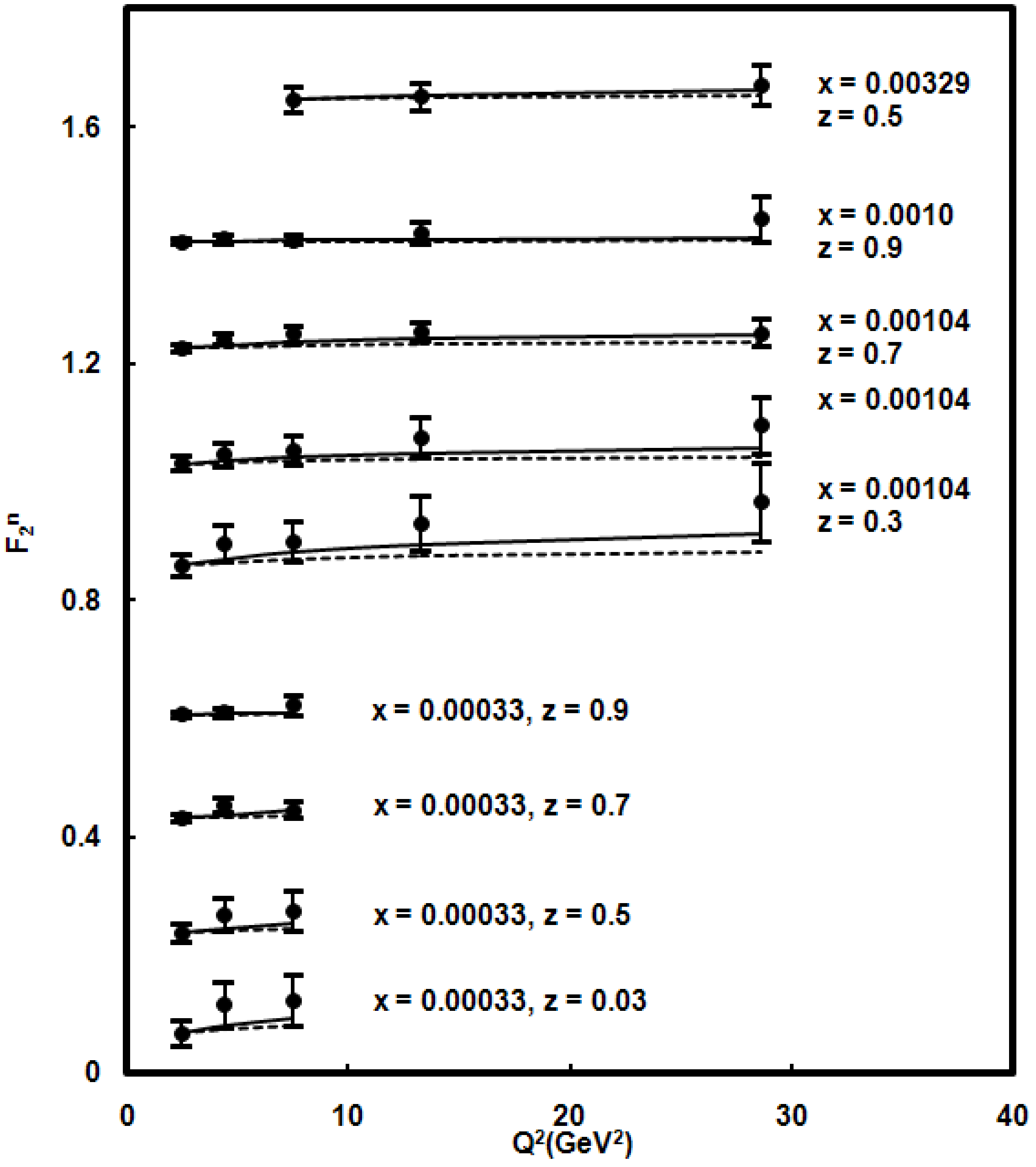

Fig- 4

In Fig.4, we present our results of t-evolutions of neutron structure functions for the representative values of $x$ given in the figures for $y = 2$ (solid lines) and $y$ maximum (dashed lines) in $\beta = \alpha^y$ relation. The dashed line also represents results of unique solution. Data points at lowest-$Q^2$ values in the figures are taken as input to test the evolution equation. Agreement with the data [10] is found to be good. NNLO results for $y = 2$ are of better agreement with experimental data in general.

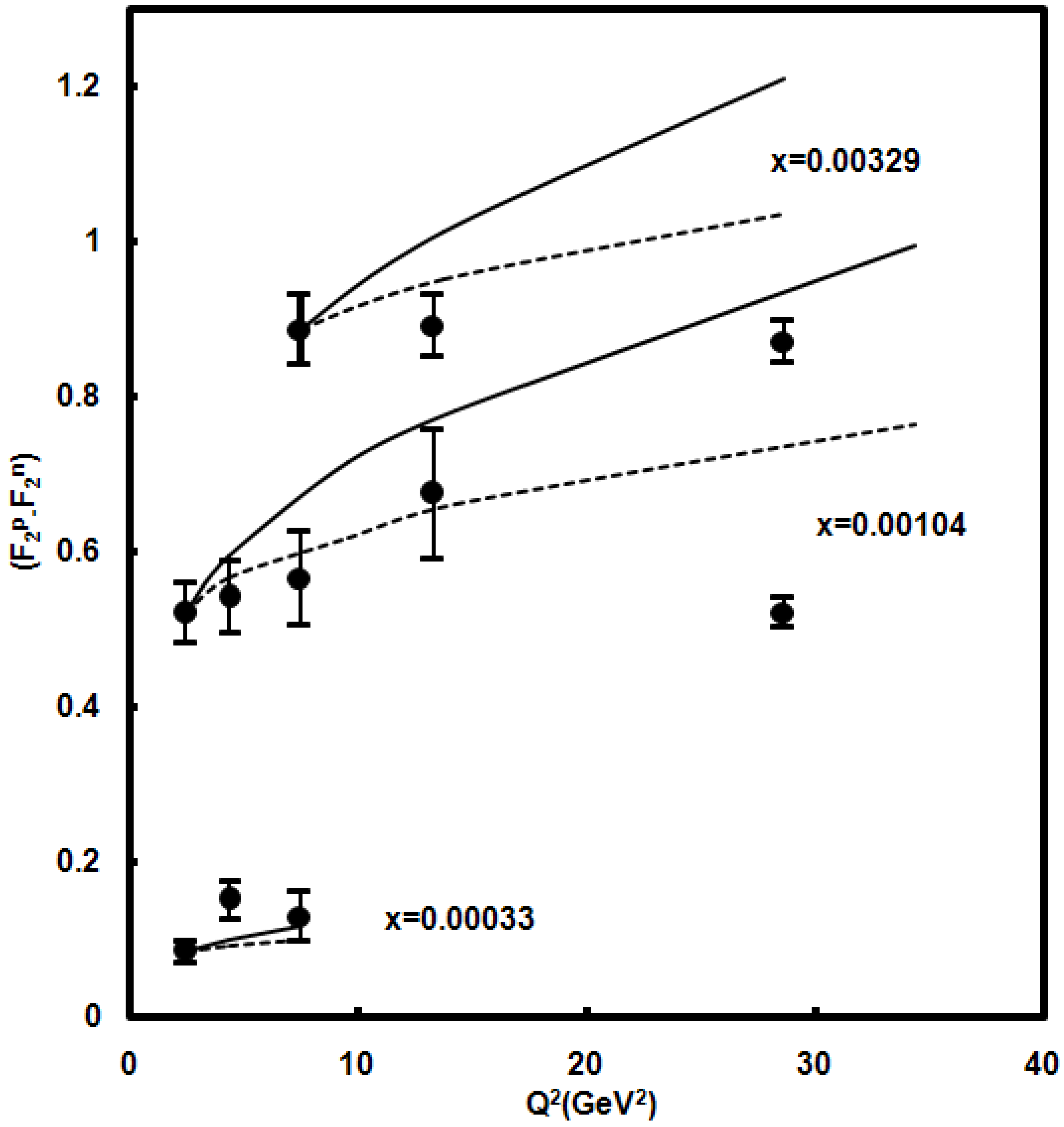

Fig-5

In Fig.5, we present our results of t-evolutions of difference of proton and neutron structure functions for the representative values of $x$ given in the figures for $y = 2$ (solid lines) and $y$ maximum (dashed lines) in $\beta = \alpha^y$ relation. The dashed line also represents results of unique solution. Data points at lowest-$Q^2$ values in the figures are taken as input to test the evolution equation. Agreement with the data [10] is found to be good. NNLO results for $y$ maximum are of better agreement with experimental data in general.

In fig.6 we present our results of $t$-evolutions of ratio of proton and neutron structure functions $F_2^p / F_2^n$ (solid lines) for the representative values of $x$ given in the figures. Though according to our theory the ratio should be independent of $t$, due to the lack of sufficient amount of data and due to large error bars, a clear cut conclusion can not be drawn.

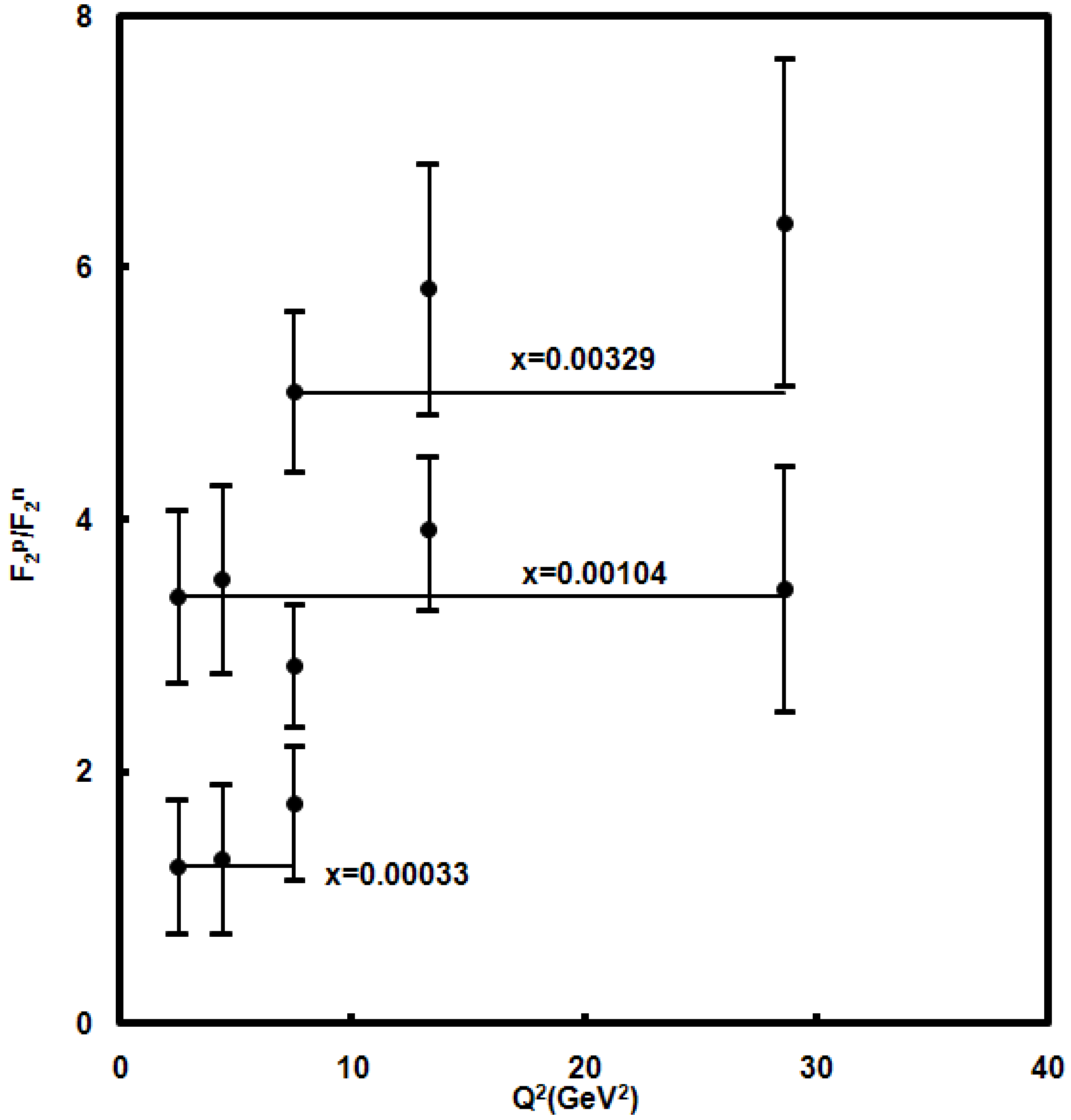

**Fig - 6**

Though we compare our results which $y = 2$ and y maximum in $\beta = \alpha^y$ relation with data, agreement of the result with experimental data is found to be excellent with $y = 2$ for $t$-evolution in next-to-next-to leading order.

In fig.7 we plot $T^2(t)$ (solid line) and $T_0T(t)$ (dotted line), $T^3(t)$ (solid line) and $T_1T(t)$ (dotted line) where $T(t) = \alpha_s(t)/2\pi$ against $Q^2$ in the $Q^2$ range $0.75 \leq Q^2 \leq 50$ GeV$^2$ . Though the explicit values of $T_1(t)$, $T_0$ are not necessary in calculating $t$-evolution of, yet we observe that for $T_1 = .0028$ and $T_0 = 0.05$, errors become minimum in the $Q^2$ range $0.5 \leq Q^2 \leq 50$ GeV$^2$.

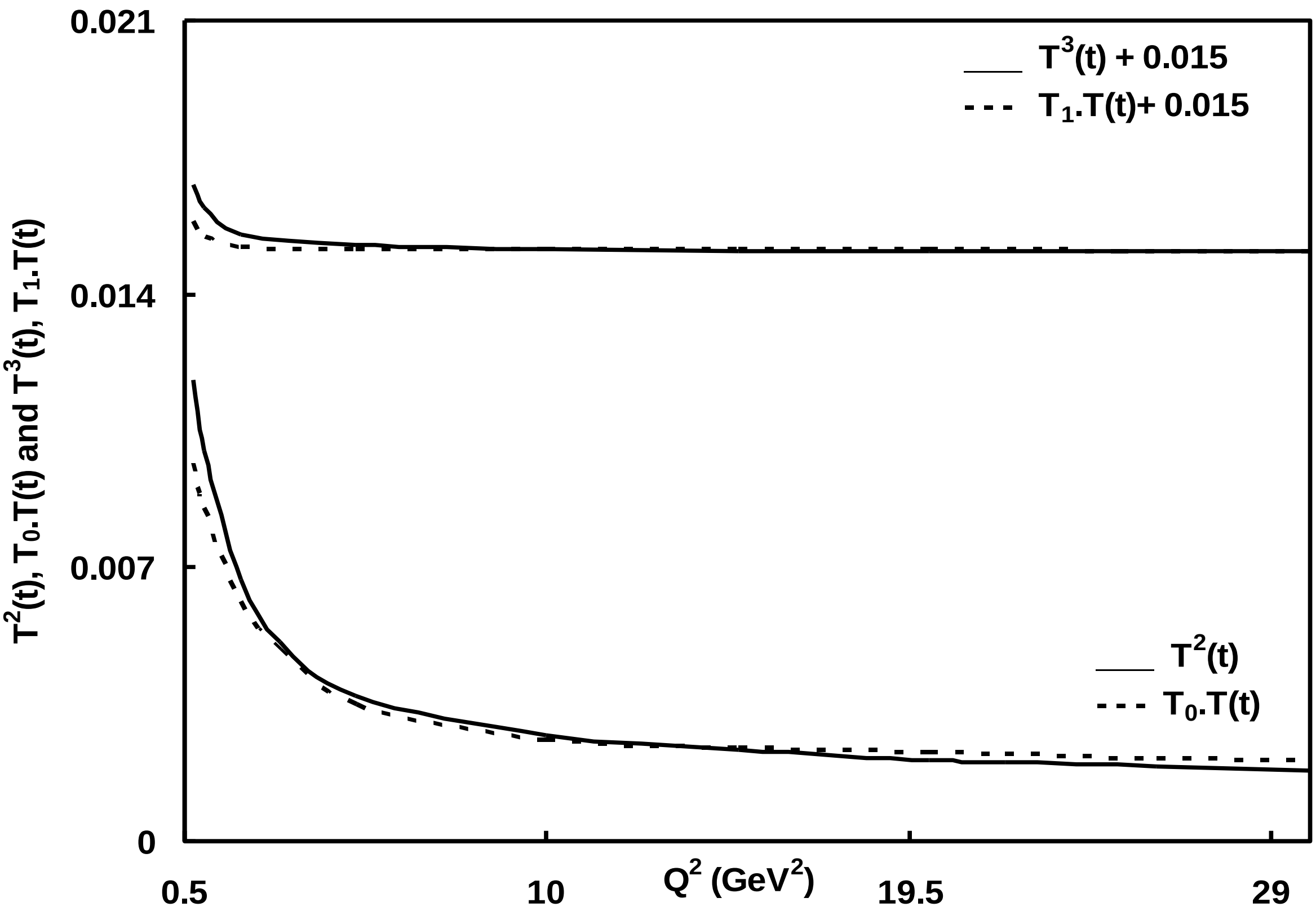

**Fig-7**

**Figure Captions**

**Fig.1:** Results of *t*-evolutions of deuteron structure functions for the representative values of *x* given in the figures for $y = 2$ (solid lines) and *y* maximum (dashed lines) in $\beta = \alpha^y$ relation. The dashed line also represents results of unique solution. For convenience, value of each data point is increased by adding $0.2i$ where $i = 0, 1, 2, 3, ...$ are the numberings of curves counting from the bottom of the lowermost curve as the 0-th order. Data points at lowest-$Q^2$ values in the figures are taken as input to test the evolution equation.

**Fig.2:** Results of *t*-evolutions of proton structure functions for the representative values of *x* given in the figures for $y = 2$ (solid lines) and *y* maximum (dashed lines) in $\beta = \alpha^y$ relation. The dashed line also represents results of unique solution. For convenience, value of each data point is increased by adding $0.2i$, where $i = 0, 1, 2, 3, ...$ are the numberings of curves counting from the bottom of the lowermost curve as the 0-th order. Data points at lowest-$Q^2$ values in the figures are taken as input to test the evolution equation.

**Fig.3:** Results of *t*-evolutions of proton structure functions $F_2^p$ with recent global paramatrization (long dashed lines) for the representative values of *x* given in the figures for $y =$

2 (solid lines) and $y$ maximum (dashed lines) in $\beta = \alpha^y$ relation. The dashed line also represents results of unique solution. Data points at lowest-$Q^2$ values in the figures are taken as input. For convenience, value of each data point is increased by adding 0.5$i$, where $i$ = 0, 1, 2, 3, ... are the numberings of curves counting from the bottom of the lowermost curve as the 0-th order.

**Fig.4:** Results of $t$-evolutions of neutron structure functions for the representative values of $x$ given in the figures for $y$ = 2 (solid lines) and $y$ maximum (dashed lines) in $\beta = \alpha^y$ relation. The dashed line also represents results of unique solution. For convenience, value of each data point is increased by adding 0.2$i$, where $i$ = 0, 1, 2, 3, ... are the numberings of curves counting from the bottom of the lowermost curve as the 0-th order. Data points at lowest-$Q^2$ values in the figures are taken as input to test the evolution equation.

**Fig.5:** Results of $t$-evolutions of difference of proton and neutron structure functions for the representative values of $x$ given in the figures for $y$ = 2 (solid lines) and $y$ maximum (dashed lines) in $\beta = \alpha^y$ relation. The dashed line also represents results of unique solution. For convenience, value of each data point is increased by adding and 0.4$i$, where $i$ = 0, 1, 2, 3, ... are the numberings of curves counting from the bottom of the lowermost curve as the 0-th order. Data points at lowest-$Q^2$ values in the figures are taken as input to test the evolution equation.

**Fig.6:** Results of $t$-evolutions of ratio of proton and neutron structure functions $F_2^p/ F_2^n$ (solid lines) for the representative values of $x$ given in the figures. Data points at lowest-$Q^2$ values in the figures are taken as input.

**Fig.7:** $T^2(t)$ (solid line) and $T_0T(t)$ (dotted line), $T^3(t)$ (solid line) and $T_1T(t)$ (dotted line),where $T(t) = \alpha_s(t)/2\pi$ against $Q^2$ in the $Q^2$ range $0.75 \leq Q^2 \leq 50$ GeV$^2$.

**Acknowledgement**

I am grateful to J K Sarma of Department of Physics of Tezpur University for the help to complete this work.